\documentclass[a4paper,11pt]{article}
\pdfoutput=1 

\usepackage{jcappub} 

\usepackage[T1]{fontenc} 
\usepackage{footnote}

\usepackage{graphicx}
\usepackage{epsfig}

\newcommand{\beq}{\begin{equation}}
\newcommand{\eq}{\end{equation}}
\newcommand{\bear}{\begin{eqnarray}}
\newcommand{\ear}{\end{eqnarray}}

\newcommand{\be}{\begin{equation}}
\newcommand{\ee}{\end{equation}}
\newcommand{\bea}{\begin{eqnarray}}
\newcommand{\eea}{\end{eqnarray}}

\begin{document}


\title{Including  birefringence into time evolution of CMB: current and future constraints}

\author{G. Gubitosi$^{1,3}$, M. Martinelli$^{2}$, L. Pagano$^{1}$}

\affiliation{$^1$Physics Department and INFN, Universit\`a di Roma ``La Sapienza'', Ple Aldo Moro 2, 00185, Rome, Italy}
\affiliation{$^2$SISSA, Via Bonomea 265, Trieste, 34136, Italy}
\affiliation{$^3$Theoretical Physics, Blackett Laboratory, Imperial College, London, SW7 2BZ, U.K.}

\abstract{
We introduce birefringence effects within the propagation history of CMB, considering the two cases of a constant effect and of an effect that increases linearly in time, as the rotation of polarization induced by birefringence accumulates during photon propagation. Both cases result into a mixing of E and B modes before lensing effects take place, thus leading to the fact that lensing is acting on spectra that are already mixed because of birefringence.
Moreover, if the polarization rotation angle increases during propagation, birefringence affects more the large scales that the small scales. We put constraints on the two cases using data from WMAP 9yr and BICEP 2013 and compare these results with the constraints obtained when the usual procedure of rotating the final power spectra is adopted, finding that this dataset combination is unable to distinguish between effects, but it nevertheless hints for a non vanishing value of the polarization rotation angle. We also forecast the sensitivity that will be obtained using data from Planck and PolarBear, highlighting how this combination is capable to rule out a vanishing birefringence angle, but still unable to distinguish the different scenarios. Nevertheless, we find that the combination of Planck and PolarBear is sensitive enough to highlight the existence of degeneracies between birefringence rotation and gravitational lensing of CMB photons, possibly leading to false detection of non standard lensing 
effects if birefringence is neglected.
}

\date{\today}
\maketitle
\section{Introduction}
\label{sec:intro}

Recent Cosmic Microwave Background (CMB) observations brought to more and more precise measurements of temperature anisotropies reaching the almost cosmic variance-limited sensitivity of Planck \cite{Ade:2013ktc,Ade:2013zuv}. While other surveys are focusing on reaching a similar sensitivity on smaller angular scales, e.g. ACT \cite{Sievers:2013ica} and SPT\cite{2011ApJ...743...28K}, other CMB experiment were designed in order to measure the CMB photons polarization properties. After the first  detection of the polarization E modes (parity-even modes) by the DASI interferometer \cite{Kovac:2002fg}, a higher sensitivity was achieved by following experiments, such as WMAP \cite{2013ApJS..208...20B}, QUIET \cite{2012ApJ...760..145Q} and BICEP\cite{Barkats:2013jfa,2014PhRvD..89f2006K,Ade:2014xna}.  Upcoming surveys are now designed to achieve even more precise measurements of E-modes and to finally detect the parity-odd modes (B modes) of CMB polarization (see e.g. ACTpol\cite{2010SPIE.7741E..1SN}, SPTpol\cite{
2012SPIE.8452E..1EA}, PolarBear\cite{Ade:2013hjl} and EBEX\cite{2014arXiv1407.6894M}). The lensing B-modes have been already detected cross-correlating a lensing template with CMB polarization maps, see e.g. \cite{Hanson:2013hsb,Ade:2014afa}, while BICEP team claimed primordial B modes detection \cite{Ade:2014xna}, although at such scales polarized dust signal must be taken into account, as pointed out in \cite{Adam:2014bub}.\\
These observations are crucial to detect signatures of the current standard cosmological model such as the B modes induced by primordial gravitational
waves and the leakage of power between E and B modes due to weak gravitational lensing of CMB photons.\\
Moreover, the precise measurement of CMB polarization allows also for tests of new physics, such as the search for CPT and Lorentz violations in the photons sector of particle physics \cite{Feng:2006dp}. 
In particular, some attention has been gained in the last few years by the search for signals of birefringence, \emph{i.e.} rotation of the photons polarization direction during \emph{in vacuo} propagation (see e.g. \cite{lue,lepora,Xia:2007qs,Kostelecky:2002hh,Finelli:2008jv,Kahniashvili:2008va,Gubitosi:2009eu,2007PhRvD..76l3014C,Balaji:2003sw, Gruppuso:2011ci, Gubitosi:2012rg,2014PhRvD..89f2006K,Giovannini:2008zv,Yadav:2012tn} and references therein), whose main effect on CMB photons consists in a mixing between E and B polarization modes. To investigate this phenomenon is crucial also because of the possible contamination that birefringence can have on primordial gravitational wave detection \cite{Zhao:2014rya,Zhao:2014yna}.\\
A similar mixing is produced on CMB polarization by weak gravitational lensing and as upcoming surveys will improve our knowledge of this effect, considering these phenomena in the right order is crucial. In fact, while CMB lensing performs its mixing at ``recent'' times, birefringence starts to take place right after recombination and it is expected to accumulate during photons propagation. Therefore the CMB spectra which are modified by lensing  do not encode only the effect of primordial anisotropies, but already contain the rotation effect due to birefringence.  However, most of previous works \cite{lepora,Xia:2007qs,Kostelecky:2002hh,Kahniashvili:2008va,Gubitosi:2009eu,2007PhRvD..76l3014C,Balaji:2003sw,Gubitosi:2012rg} apply the rotation due to birefringence on the lensed CMB spectra. This procedure is correct only when one expects the polarization rotation not to be a genuine physical effect but to be  due to miscalibration of the polarimeters \cite{malapola,Kaufman:2014rpa}.\\
In this paper we address this issue comparing the results obtained with currently available datasets in both the early and late time rotation cases, using WMAP and the more polarization-oriented survey BICEP. We also inquire about the possibility of future CMB surveys to detect a non-zero birefringence effect or rather to rule it out exploiting forecasted datasets and we investigate the possibility of future surveys to distinguish among different types of polarization rotation. \\
Furthermore, as birefringence modifies also the power spectrum of $B$ modes, giving it an additional contribution due to the leakage from the $E$ modes, we also investigate the possible degeneracies between birefringence parameters and CMB lensing. Indeed lensing effect on CMB spectra, parametrized by the lensing amplitude $A_L$ \cite{Calabrese:2008rt}, also leads to a leakage from $E$ to $B$ modes, so neglecting the presence of birefringence can in principle produce a misleading detection of a non standard lensing effect ($A_L\neq1$).

The paper is organized as follows. In Section \ref{sec:theo} we briefly review birefringence theory and its motivations, also describing how it affects CMB power spectra. In Section \ref{sec:ana} we describe the performed analysis and the datasets used to constrain birefringence parameters. General results are presented in Section \ref{sec:res} while we discuss them in the concluding Section \ref{sec:conc}.\\

\section{Theory of birefringence}
\label{sec:theo}

Birefringence is the rotation of linear polarization direction during the propagation of radiation in vacuo.

The standard way birefringence is formalized in the literature is through a  sudden rotation of the polarization after photon propagation from the last scattering surface to now. 
There are however exceptions to this, see for example \cite{Finelli:2008jv, Liu:2006uh}, where the amount of birefringence depends on the evolution of a cosmological scalar
 field (see also \cite{Gubitosi:2009eu} for a discussion on the accuracy of this "sudden rotation" approximation).

If the polarization direction rotates counterclockwise (looking at the source) of an angle $\beta>0$, then the Stokes parameters  $Q$ and $U$ get mixed in the following way\footnote{The Stokes parameters are defined in the standard frame used for CMB, see \cite{Hinshaw:2008kr, Zaldarriaga:1996xe}, so that a counterclockwise rotation of  the polarization direction (looking at the photons coming toward us) corresponds to a rotation of the reference frame from the $\hat x$ axis to the $\hat y$ axis.}:
\bea
Q&=&\tilde Q \cos 2\beta+\tilde U \sin 2\beta\nonumber\\
U&=&\tilde U \cos 2\beta-\tilde Q\sin 2\beta \label{eq:StokesMixing}
\eea
 and as a consequence the power spectra become\footnote{For reasons that will be clear later one usually rotates only the spectra at multipole $\ell \gtrsim 20$. The lower multipoles are not rotated \cite{wmap5}.}:
 \begin{eqnarray}
 C_\ell^{EE}&=&\tilde C_\ell^{EE}  \cos^2\left(2 \beta\right) +\tilde C_\ell^{BB}  \sin^2\left(2 \beta\right)  -\tilde C_{\ell}^{EB}\sin \left(4\beta\right)\nonumber \\
C_\ell^{BB}&=&\tilde C_\ell^{EE}  \sin^2\left(2 \beta\right)  +\tilde C_\ell^{BB}  \cos^2\left(2 \beta\right) +\tilde C_\ell^{EB}  \sin\left(4 \beta\right) \nonumber \\
C_\ell^{EB}&=&\frac{1}{2}\left(\tilde C_\ell^{EE}-\tilde C_\ell^{BB}\right)  \sin\left(4 \beta\right)   +\tilde C_\ell^{EB}\left(\cos^{2}\left(2\beta\right)-  \sin^2\left(2 \beta\right)  \right) \nonumber \\
C_\ell^{TE}&=&\tilde C_\ell^{TE}  \cos\left(2 \beta\right) -\tilde C_\ell^{TB}  \sin\left(2 \beta\right)     \nonumber \\
C_\ell^{TB}&=&\tilde C_\ell^{TE}  \sin\left(2 \beta\right)   +\tilde C_\ell^{TB}  \cos\left(2 \beta\right)    \label{eq:powerspectrareloaded}
\end{eqnarray}
 The $\tilde C_{\ell}$ are the spectra in absence of polarization rotation (no birefringence), while the $C_{\ell}$ are the observed spectra and we allowed for the presence of non zero parity-violating cross-correlation spectra before the rotation occurs, i.e. $\tilde{C}_\ell^{EB}$ and $\tilde{C}_\ell^{TB}$.

This way of treating polarization rotation is exact when considering the modification of the spectra that one would expect from a systematic miscalibration of the polarimeters \cite{malapola,Kaufman:2014rpa}, as in this case one would have a genuine effect on the final spectra.
However, when dealing with birefringence as the effect of some new physics phenomenon, eqs. (\ref{eq:powerspectrareloaded}) can only be considered as an approximation. In fact birefringence is a phenomenon due to anomalous photon propagation \cite{Carroll:1989vb,Liu:2006uh,Finelli:2008jv,Myers:2003fd,Kahniashvili:2008va,Gubitosi:2009eu}, that accumulates during propagation from last scattering  to now. 
In this case the amount of rotation is time dependent, given by $\alpha(\eta)$ as a function of conformal time from last scattering $\eta$. Depending on the model considered, the actual form of the time dependence of the amount of rotation can vary.

For a time-dependent rotation of polarization direction, equation (\ref{eq:StokesMixing}) is easily generalized:
\bea
Q(\eta)&=&\tilde Q(\eta) \cos 2\alpha(\eta)+\tilde U(\eta) \sin 2\alpha(\eta)\nonumber\\
U(\eta)&=&\tilde U(\eta) \cos 2\alpha(\eta)-\tilde Q(\eta)\sin 2\alpha(\eta) \label{eq:StokesMixingAlpha}
\eea

This induces a modification of the Boltzmann equation for the evolution of polarization perturbations, $\Delta_{Q\pm i U}(\vec k,\eta)$. In Fourier space \cite{Kosowsky:1996yc}: 
\bea
\dot \Delta_{Q\pm i U}(\vec k,\eta)+i k \mu \Delta_{Q\pm i U}(\vec k,\eta)&=&\dot \tau (\eta)\left[ -\Delta_{Q\pm i U}(\vec k,\eta)- \sum_{m} \sqrt{\frac{6\pi}{5}}    \,_{\pm2}Y_{2}^{m}(\hat n) S_{p}^{m}(\vec k,\eta)  \right]\nonumber\\
&& \mp i 2 \dot \alpha(\eta)  \Delta_{Q\pm i U}(\vec k,\eta)
\eea 
where  derivatives are taken with respect to conformal time, $\mu$ is the cosine of the angle between the photon  propagation direction and the Fourier mode $\vec k$. $\dot \tau(\eta)$ is the differential optical depth, $\dot \tau(\eta)\equiv n_{e} \sigma_{T} a(\eta)$, where $n_{e}$ and $\sigma_{T}$ are, respectively, the free electron number density and the Thomson cross section, and $a$ is the scale factor. $\,_{\pm 2}Y^{m}_{2}$ are spin-weighted spherical harmonics with spin $\pm 2$ and $S_{P}^{(m)}$ is the polarization source ($m=0,\pm1,\pm2$ indicates, respectively, scalar, vector and tensor perturbations ). The last term in the equation is the one due to birefringence, and its form can be easily deduced by taking a time derivative of the appropriate combination of eqs. (\ref{eq:StokesMixingAlpha}).

To formally integrate over the line of sight one observes that 
\bea
\dot \Delta_{Q\pm i U}+(i k \mu +\dot \tau\pm i 2\dot \alpha)\Delta_{Q\pm i U}=e^{-i k \mu\eta}e^{\tau(\eta)}e^{\mp i2 \alpha(\eta)}\frac{d}{d\eta}\left[e^{i k \mu\eta}e^{-\tau(\eta)}e^{\pm i2 \alpha(\eta)} \Delta_{Q\pm i U}\right].
\eea
where we have defined the  total optical depth $\tau(\eta)\equiv \int_{\eta}^{\eta^{*}} \dot \tau(\eta') d\eta'$, with $\eta^{*}$ the time at recombination, such that $d\tau/d\eta=-\dot \tau$. The total amount of polarization rotation after propagation for a time $\eta$ from recombination is $\alpha(\eta)=\int_{\eta^{*}}^{\eta} \dot \alpha(\eta')d\eta'$.
The integration along the line of sight then gives:
\be
\Delta_{Q\pm i U}(\eta_{0})=\int_{0}^{\eta_{0}}d\eta\, e^{i k \mu(\eta-\eta_{0})}e^{-[\tau(\eta)-\tau(\eta_{0})]}e^{\pm i2 \left[\alpha(\eta)-\alpha(\eta_{0})\right]} \dot \tau (\eta) \sum_{m} \sqrt{\frac{6\pi}{5}}    \,_{\pm2}Y_{2}^{m}(\hat n) S_{p}^{m}(\vec k,\eta) 
\ee
To go to $E,B$ space one exploits the relations \cite{Zaldarriaga:1996xe}:
\bea
\Delta_{E}&=& -\frac{1}{2}\left(\bar{\text{\dh}}^{2} \Delta_{Q+ iU}+{\text{\dh}}^{2} \Delta_{Q- iU}\right)\\
\Delta_{B}&=& -\frac{i}{2}\left(\bar{\text{\dh}}^{2} \Delta_{Q+ iU}-{\text{\dh}}^{2} \Delta_{Q- iU}\right)
\eea
Comparing with \cite{Zaldarriaga:1996xe}, eqs. (12)-(15) for  scalar perturbations and eqs. (26)-(30) for tensor perturbations, it is easy to follow the same procedure outlined there and get: 
\bea\label{eq:perturbations1}
\Delta_{E,\ell}^{(S)}(k,\eta_{0})&=&\sqrt{\frac{\ell+2}{\ell-2}}\int_{0}^{\eta_{0}}d\eta S_{E}^{(S)}(k,\eta)  \cos{\left(2\,\delta\alpha(\eta)\right)}
\ j_{\ell}(k(\eta_{0}-\eta))\\
\Delta_{B,\ell}^{(S)}(k,\eta_{0})&=&\sqrt{\frac{\ell+2}{\ell-2}}\int_{0}^{\eta_{0}}d\eta S_{E}^{(S)}(k,\eta)\sin{\left(2\,\delta\alpha(\eta)\right)}
 j_{\ell}(k(\eta_{0}-\eta))\\
\Delta_{E,\ell}^{(T)}(k,\eta_{0})&=&\int_{0}^{\eta_{0}}d\eta \left[S_{E}^{(T)}(k,\eta)  \cos{\left(2\,\delta\alpha(\eta)\right)}-S_{B}^{(T)}(k,\eta)\sin{\left(2\,\delta\alpha(\eta)\right)}
\right] j_{\ell}(k(\eta_{0}-\eta))\nonumber\\
 &&\\
\Delta_{B,\ell}^{(T)}(k,\eta_{0})&=&\int_{0}^{\eta_{0}}d\eta \left[S_{B}^{(T)}(k,\eta)  \cos{\left(2\,\delta\alpha(\eta)\right)}+S_{E}^{(T)}(k,\eta)\sin{\left(2\,\delta\alpha(\eta)\right)}
\right] j_{\ell}(k(\eta_{0}-\eta))\nonumber\\
&&\label{eq:perturbations2}
\eea
where $S_{E}^{(S,T)}$ and $S_{B}^{(T)}$ are the sources for $E$ and $B$ modes respectively as they appear in eqs. (18) and (30) of  \cite{Zaldarriaga:1996xe} and $\delta\alpha(\eta)\equiv \alpha(\eta)-\alpha(\eta_{0}) = \int_{\eta_{0}}^{\eta} \dot \alpha(\eta')d\eta'$. 

The power spectra are  computed in the standard way as:
\be
C_{\ell}^{XY}=(4\pi)^{2} \int d k k^{2} P_{\phi}(k) \Delta_{X,\ell}(k,\eta_{0}) \Delta_{Y,\ell}^{*}(k,\eta_{0})\label{eq:powerspectranew}
\ee
 where $ P_{\phi}(k)$ is the initial power spectrum and  $\Delta_{X,\ell}(k,\eta_{0})$ is the perturbation of the mode $X=\{T,E,B\}$ in Fourier space at time $\eta_{0}$ (the one for temperature is standard and can be found in \cite{Zaldarriaga:1996xe}, the ones for $E$ and $B$ modes are given above). \\
In the following we will focus on a linear time dependence, parameterized as 
\be\alpha(\eta)=\alpha_{1} \frac{\eta}{\eta_{0}}.\label{eq:LinearAlpha}\ee 
In this case $\delta\alpha(\eta)= \alpha_{1} \frac{\eta}{\eta_{0}}-\alpha_{1}$.
Motivations for studying this particular time dependence come from some quantum-gravity-motivated studies \cite{Myers:2003fd, Kahniashvili:2008va, Gubitosi:2009eu, Gubitosi:2012rg}, where the amount of rotation is quadratically dependent on the energy of the photons and linearly dependent on the propagation time.
One might of course consider more complicated functional forms for the time dependence, for example one could link birefringence to a coupling of photons to quintessence fields \cite{Liu:2006uh} or to pseudo-scalar fields \cite{Finelli:2008jv}. In that case the time evolution of the birefringence effect is linked to the time evolution of the fields.

\begin{figure}[!htb]
\begin{center}
\hspace*{-1cm}
\begin{tabular}{cc}
\includegraphics[width=18cm]{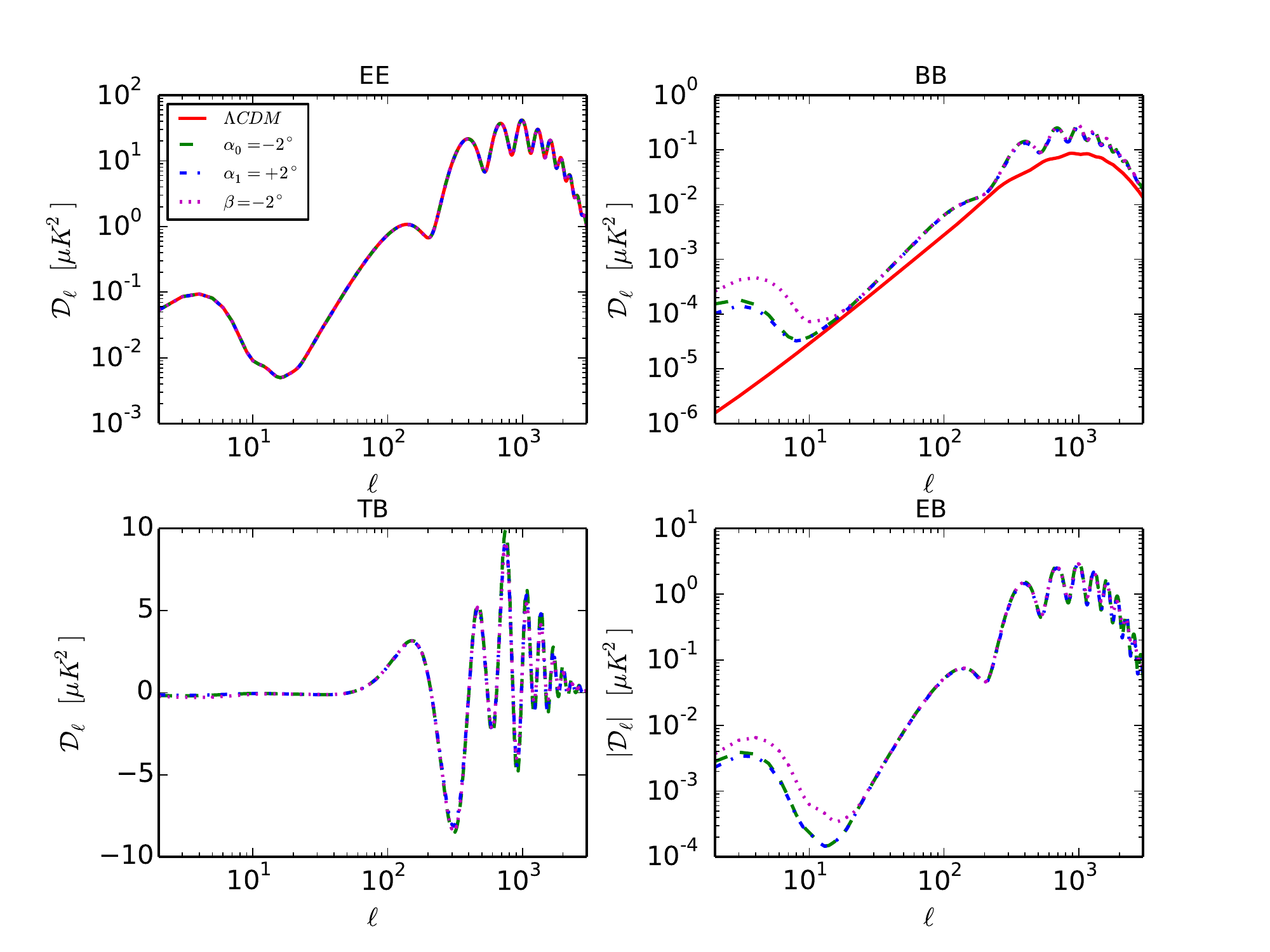}
 \end{tabular}
\caption{CMB power spectra obtained in standard $\Lambda$CDM (red solid line), late time constant rotation (pink dotted line), early time constant rotation (green dashed line) and time evolving rotation (blue dashed line) cosmologies.}
\label{fig:spectra}
\end{center}
\end{figure}

In Figure \ref{fig:spectra} we compare the polarization power spectra that one expects in standard $\Lambda$CDM model, with the ones expected if a polarization rotation is present, taking into account different possibilities: rotation that acts on the time-evolved spectra (\emph{i.e.} the one described by the parameter $\beta$ in  eqs. (\ref{eq:powerspectrareloaded})), rotation that evolves with photon propagation with a linear time dependence (described by eqs. (\ref{eq:perturbations1})-(\ref{eq:powerspectranew}), with relevant parameter $\alpha_1$ given in (\ref{eq:LinearAlpha})), a constant rotation that acts on the time evolved spectra, before lensing (see next subsection for details) described by the parameter $\alpha_0$. 
 Note that in all of the three cases the effects of lensing are present, even though they are treated differently, as is explained in the following subsection. In particular, in the time-evolving case (parameter $\alpha_1$), lensing acts on the spectra that were rotated by the time-evolving birefringence effect.

\begin{figure}[!h]
\begin{center}
\hspace*{-1cm}
\begin{tabular}{cc}
 \includegraphics[width=12cm]{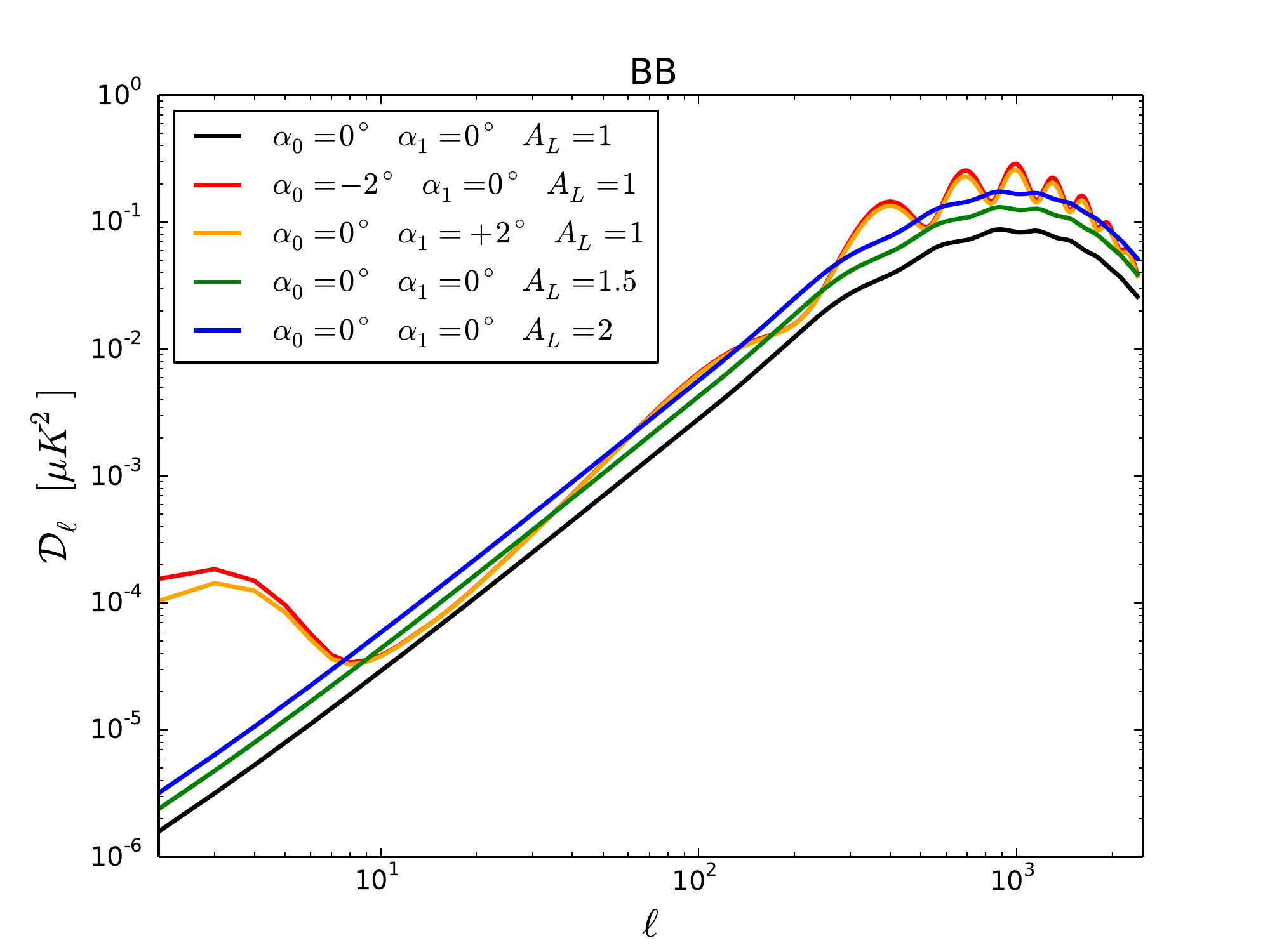}
 \end{tabular}
  \caption{BB spectra for $\Lambda$CDM (black line), rotation cosmologies with a non zero $\alpha_1$ (orange line) and $\alpha_0$ (red line), compared with the spectra produced by cosmologies without rotation, but with $A_L=1.5$ (green line) and $A_L=2$ (blue line).}
\label{fig:lensdeg}
\end{center}
\end{figure}

Figure \ref{fig:spectra} shows how the introduction of the polarization rotation angle transfers power from the $EE$ spectrum to the other polarization spectra both producing $C_\ell^{TB}$ and $C_\ell^{EB}$, which are vanishing in a standard $\Lambda$CDM framework where no rotation is present, and contributing to the $C_\ell^{BB}$ spectrum.\\
The latter effect is of particular interests as another transfer of power from $C_\ell^{EE}$ to $C_\ell^{BB}$, acting on the same range of angular scales, is produced by gravitational lensing of CMB photons. The impact of the lensing effect on B mode spectrum can be parametrized through the lensing amplitude $A_L$ which is equal to $1$ in a standard $\Lambda$CDM cosmology; this leads, as can be seen in Figure \ref{fig:lensdeg}, to possible degeneracies between lensing and polarization rotation, as the effects brought by birefringence can be (partially) mimicked by an enhanced lensing amplitude. Therefore it is crucial to treat in a proper way the combination of the two physical effects on CMB spectra, as it is done in the following section.\\

\subsection{Effects of lensing}

As we already mentioned, the way birefringence is usually treated is through the rotation of the 'final' power spectra, resulting from propagation of radiation from the last scattering surface until now. This means that birefringence rotates spectra that have already been mixed by lensing \footnote{  Exceptions to this treatment of lensing are found in \cite{Finelli:2008jv, Liu:2006uh}, where the amount of birefringence depends on the evolution of a cosmological scalar field. However, a detailed study of the degeneracy between lensing and birefringence is missing in those papers.}.

It should be clear now that this procedure is consistent only if one assumes that the spectra rotation is due to some systematic effect, which acts at the level of the detectors.

If birefringence is due to genuinely physical effects acting along photon propagation, then one should take into account the fact that lensing will mix spectra that have already been rotated by birefringence. 
In particular birefringence will have already generated B modes from E modes, so that, for example, $C_{\ell}^{TB}$ and $C_{\ell}^{EB}$ are non zero.
This means that the formulae for the lensed spectra have to be derived in the most general case when all the power spectra are non zero. This is shown in the following, concentrating only on the quantities that involve the polarization field, since the temperature field is not affected by birefringence.

The construction of lensed spectra, which we will call $\bar C_\ell$, relies on the real-space correlation functions \cite{Chon:2003gx}
\bea
\xi_X(\gamma)&\equiv &<T(\hat n_1)P(\hat n_2)>\nonumber\\
\xi_+(\gamma)&\equiv &<P^*(\hat n_1)P(\hat n_2)>\nonumber\\
\xi_-(\gamma)&\equiv &<P(\hat n_1)P(\hat n_2)>
\eea
where $\gamma$ is the angle between directions $\hat n_1$ and $\hat n_2$, $P=Q+i U$ is the polarization field defined in the local basis with $\hat x$ direction along the geodesics between $\hat n_1$ and $\hat n_2$.

It can be shown (see \cite{Chon:2003gx}) that in absence of lensing these correlation functions are related to the power spectra and to the geometrical factors $d_{ss'}^\ell(\gamma)\equiv \sum_m \,_sY_{\ell m}^*(\hat n_1) \,_{s'}Y_{\ell m}(\hat n_2)$ in the following way:
\bea
\xi_X(\gamma)&=& \sum_\ell \frac{2\ell +1}{4\pi}(C_\ell^{TE}-i C_\ell^{TB})d_{20}^\ell(\gamma) \nonumber\\
\xi_+(\gamma)&=& \sum_\ell \frac{2\ell +1}{4\pi}(C_\ell^{EE}+ C_\ell^{BB})d_{22}^\ell(\gamma) \nonumber\\
\xi_-(\gamma)&=& \sum_\ell \frac{2\ell +1}{4\pi}(C_\ell^{EE}-C_\ell^{BB}-2i C_\ell^{EB})d_{2-2}^\ell(\gamma)  \label{eq:corrFunctionsCl}
\eea
Note that we already are considering the possibility of having the parity-violating spectra $C_\ell^{EB}$ and $C_\ell^{TB}$.

When lensing is introduced the above formulas have to be modified introducing terms related to the power spectrum of the lensing potential. A detailed computation which does not consider the parity violating spectra can be found in \cite{Challinor:2005jy}. When including the parity violating spectra $C_\ell^{EB}$ and $C_\ell^{TB}$ the relevant formulas found in \cite{Challinor:2005jy} generalize to:
\bea
\bar\xi_X(\gamma)&=& \sum_{\ell}  \frac{2\ell +1}{4\pi} (C_\ell^{TE}-i C_\ell^{TB})\Big\{d_{20}^\ell X_{022}X_{000}+C_{gl,2}\frac{2 X_{000}'}{\sqrt{\ell(\ell+1)}}(X_{112}d_{11}^\ell+X_{132} d_{3-1}^\ell)\nonumber\\
&&\qquad +\frac{1}{2} C_{gl,2}[d_{20}^\ell(2 X_{022}'X_{000}'+X_{220}^2)+d_{-24}^\ell X_{220}X_{242}] \Big\}\nonumber\\
\bar\xi_+(\gamma)&=& \sum_\ell \frac{2\ell +1}{4\pi}(C_\ell^{EE}+ C_\ell^{BB})\Big\{d_{22}^\ell X_{022}^2+2 C_{gl,2} X_{132}X_{121}d_{31}^\ell +C_{gl,2}^2[d_{22}^\ell (X_{022}')^2\nonumber\\
&&\qquad +d_{40}^\ell X_{220}X_{242}] \Big\} \nonumber\\
\bar\xi_-(\gamma)&=& \sum_\ell \frac{2\ell +1}{4\pi}(C_\ell^{EE}-C_\ell^{BB}-2i C_\ell^{EB})\Big\{d_{2-2}^\ell X_{022}^2+C_{gl,2} [X_{121}^2d_{1-1}^\ell+X_{132}^2d_{3-3}^\ell] \nonumber\\
&&\qquad +\frac{1}{2}C_{gl,2}^2[ 2 d_{2-2}^\ell (X_{022}')^2+d_{00}^\ell X_{220}^2+d_{4-4}^\ell X_{242}^2] \Big\} \label{eq:LensedXi}
\eea
The $\bar \xi_{\{X,+,-\}}$ are the lensed correlation functions, $X_{ijk}$ and $C_{gl,2}$ depend on the lensing potential and geometrical factors and are defined in \cite{Challinor:2005jy}. Note that the power spectra appearing here are the non-lensed ones, but \emph{include already the effect of birefringence}  if one is not considering the late-time birefringence case.

Once the lensed correlation functions have been computed as in the above equations, the lensed spectra $\bar C_\ell$ can be derived by using the relations (\ref{eq:corrFunctionsCl}) and the orthogonality of the $d_{ss'}^\ell$: 
\bea
\bar C_\ell^{TE}-i \bar C_\ell^{TB}=2\pi \int_{-1}^1 \bar \xi_X d_{20}^\ell d\cos \gamma\nonumber\\
\bar C_\ell^{EE}+ \bar C_\ell^{BB}=2\pi \int_{-1}^1 \bar \xi_+ d_{22}^\ell d\cos \gamma\nonumber\\
\bar C_\ell^{EE}-\bar C_\ell^{BB}-2 i \bar C_\ell^{EB}=2\pi \int_{-1}^1 \bar \xi_- d_{2-2}^\ell d\cos \gamma \label{eq:Clfromxi}
\eea

Comparing equations (\ref{eq:LensedXi}) and (\ref{eq:Clfromxi}) one sees that  a common effect of lensing and birefringence is the generation of B modes from E modes. So it is interesting to investigate  the degeneracy between the two effects.

A simple way to do this is by studying the difference between a 'late-time' birefringence and an 'early-type' birefringence. In both cases one considers a constant polarization rotation angle, but while in the first case rotation acts on lensed spectra, which are evaluated in the standard way, in the second one birefringence acts on the spectra propagated from last scattering surface, but \emph{before} lensing is applied.

More in detail, in the late-time birefringence case the observed spectra are given by  
 \begin{eqnarray}
 C_\ell^{EE}&=&\bar C_\ell^{EE}  \cos^2\left(2 {\beta}\right) +\bar C_\ell^{BB}  \sin^2\left(2 {\beta}\right)  -\bar C_{\ell}^{EB}\sin \left(4{\beta}\right)\nonumber \\
C_\ell^{BB}&=&\bar C_\ell^{EE}  \sin^2\left(2 {\beta}\right)  +\bar C_\ell^{BB}  \cos^2\left(2 {\beta}\right) +\bar C_\ell^{EB}  \sin\left(4 {\beta}\right) \nonumber \\
C_\ell^{EB}&=&\frac{1}{2}\left(\bar C_\ell^{EE}-\bar C_\ell^{BB}\right)  \sin\left(4 {\beta}\right)   +\bar C_\ell^{EB}\left(\cos^{2}\left(2{\beta}\right)-  \sin^2\left(2 {\beta}\right)  \right) \nonumber \\
C_\ell^{TE}&=&\bar C_\ell^{TE}  \cos\left(2 {\beta}\right) -\bar C_\ell^{TB}  \sin\left(2 {\beta}\right)     \nonumber \\
C_\ell^{TB}&=&\bar C_\ell^{TE}  \sin\left(2 {\beta}\right)   +\bar C_\ell^{TB}  \cos\left(2 {\beta}\right)    \label{eq:powerspectrareloaded}
\end{eqnarray}
 using as $\bar C_\ell$ the non-rotated, but lensed, spectra, computed in the standard way (see e.g. \cite{Challinor:2005jy}).
In the early-time birefringence case, one applies the procedure described in this subsection, using in eq. (\ref{eq:LensedXi}) the spectra obtained through:
 \begin{eqnarray}
 C_\ell^{EE}&=&\tilde C_\ell^{EE}  \cos^2\left(2 {\alpha_0}\right) +\tilde C_\ell^{BB}  \sin^2\left(2 {\alpha_0}\right)  -\tilde C_{\ell}^{EB}\sin \left(4{\alpha_0}\right)\nonumber \\
C_\ell^{BB}&=&\tilde C_\ell^{EE}  \sin^2\left(2 {\alpha_0}\right)  +\tilde C_\ell^{BB}  \cos^2\left(2 {\alpha_0}\right) +\tilde C_\ell^{EB}  \sin\left(4 {\alpha_0}\right) \nonumber \\
C_\ell^{EB}&=&\frac{1}{2}\left(\tilde C_\ell^{EE}-\tilde C_\ell^{BB}\right)  \sin\left(4 {\alpha_0}\right)   +\tilde C_\ell^{EB}\left(\cos^{2}\left(2{\alpha_0}\right)-  \sin^2\left(2 {\alpha_0}\right)  \right) \nonumber \\
C_\ell^{TE}&=&\tilde C_\ell^{TE}  \cos\left(2 {\alpha_0}\right) -\tilde C_\ell^{TB}  \sin\left(2 {\alpha_0}\right)     \nonumber \\
C_\ell^{TB}&=&\tilde C_\ell^{TE}  \sin\left(2 {\alpha_0}\right)   +\tilde C_\ell^{TB}  \cos\left(2 {\alpha_0}\right)    \label{eq:powerspectrareloaded}
\end{eqnarray}
 where $\tilde C_\ell$ are now the spectra one obtains after propagation of photons from last scattering to now, without the inclusion of lensing effect.
 
The difference between the spectra obtained with a late-time birefringence and an early-time birefringence can be seen in Fig.\ref{fig:spectra}.

\section{Analysis}\label{ana}
\label{sec:ana}

The baseline set of cosmological parameters we evaluate is composed by the standard ones, namely the baryon and CDM physical matter
density $\Omega_{b}h^2$ and $\Omega_{c}h^2$, the scalar spectral index $n_s$, the optical depth $\tau$, the scalar amplitude as evaluated at a pivot scale $k = 0.002 Mpc^{-1}$ $A_s$ the angular size of the sound horizon at last scattering surface $\theta$. To this set of parameters, we add the effect of the two constant rotation models discussed above, parametrized by $\alpha_0$ (early times rotation) and $\beta$ (late times rotation) and the parameter describing the time-varying rotation $\alpha_1$. We fit the $C_\ell$ obtained through the theory discussed above combining datasets from WMAP latest release \cite{2013ApJS..208...20B} and BICEP 2013 release \cite{Barkats:2013jfa}. We do not exploit the latest release of BICEP \cite{Ade:2014xna} experiment as in that case the TB and EB spectra are used to calibrate the survey to achieve a vanishing rotation angle.  As done in \cite{wmap5} in the late times rotation (i.e. $\beta$) we do not rotate multipoles below $\ell=23$, because the polarization signal, 
at those multipoles, was generated during reionization and so we would measure only the polarization rotation between the reionization and present epoch \cite{Liu:2006uh,malapola}. Therefore, one would in principle need to separately analyze the two multipole regions (below and above $\ell=23$), but the low-$\ell$ one has a much poorer constraining power \cite{wmap5}.\\
We assume flat priors on the sampled parameters and we exploit MCMC technique through the publicly available package \texttt{cosmomc} \cite{Lewis:2002ah} with a convergence diagnostic using the Gelman and Rubin statistics.\\
Furthermore, once the best fit values for the WMAP+BICEP are obtained, we use these as the fiducial cosmologies to build forecasted datasets for Planck \cite{planck} and PolarBear \cite{Ade:2013hjl} data, in order to investigate how the upcoming data from these  surveys will tighten the previously obtained constraints and if they would be able to distinguish the effects of the different rotation parameters. We produce two sets of forecasted datasets, one assuming a cosmology with the WMAP+BICEP best fit for $\alpha_0$ and one with the best fit for $\alpha_1$. The first set is analyzed with a varying $\alpha_0$ and with a varying $\beta$ in order to understand if the combination of Planck and PolarBear will be able to distinguish between an early constant rotation and a late constant one. The second set is instead analyzed once with a varying $\alpha_1$ and once with a varying $\alpha_0$ allowing to inquire about the different effects of a time evolving and a constant rotation. Both these analysis are 
performed by probing the standard set of parameters alongside those related to rotation effects and using the same MCMC technique employed for WMAP and BICEP. The two datasets are also analyzed with the assumption of no rotation (thus allowing only standard $\Lambda CDM$ parameters to vary) in order to investigate whether the assumption of no rotation would lead to a bias in the recovered best fit of standard parameters in the eventuality of a non vanishing birefringence effect.\\
Finally, we also use the two combinations of simulated datasets to inquire about the possible degeneracy between weak gravitational lensing of the CMB and rotation of the spectra introduced by the birefringence effect. This is done by analyzing the datasets not varying any rotation parameter, but allowing the lensing amplitude $A_L$  to vary. As stated in Section \ref{sec:theo} this parameter enhances the effect of lensing on CMB spectra if raised above the standard $\Lambda$CDM value $A_L=1$, thus if the model $\Lambda$CDM$+A_L$ is used to fit the simulated dataset, a non standard value of $A_L$ could be detected in order to mimic the enhancement of $BB$ modes produced by birefringence rotation angles.

\section{Results}
\label{sec:res}

\begin{figure}[!h]
\begin{center}
\hspace*{-1.5cm}
\begin{tabular}{cc}
\includegraphics[width=14cm]{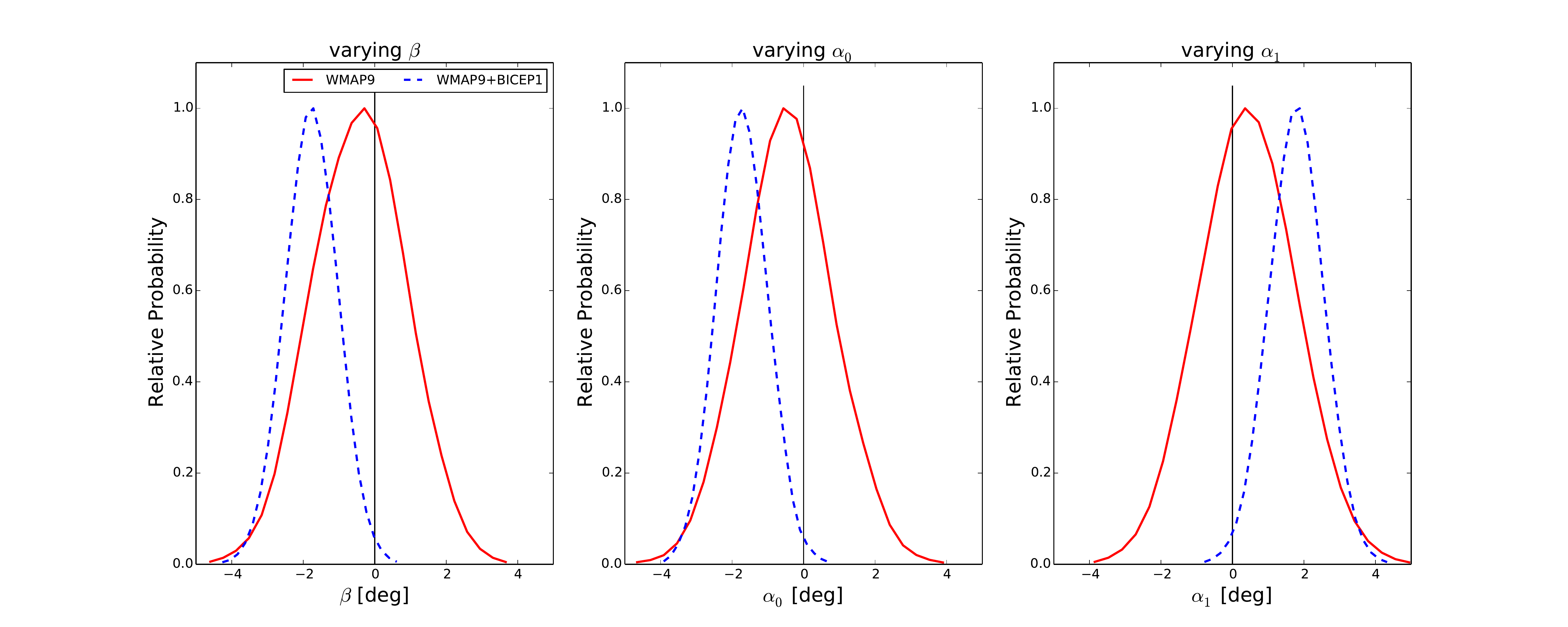}
 \end{tabular}
\caption{Posterior distributions for (from right to left) $\alpha_1$, $\alpha_0$ and $\beta$  using WMAP (red continuous lines) and WMAP+BICEP (blue dashed lines).}
\label{fig:monodim}
\end{center}
\end{figure}

As stated in the previous section, we exploit currently available data from WMAP and BICEP in order to constrain the amplitude of the birefringence parameter, both in the time evolving ($\alpha_1$) and constant ($\alpha_0$) case, and of the late time rotation angle ($\beta$). In Table \ref{tab:resWMAP} the results obtained with this experimental configuration are reported and it is possible to notice how the standard cosmological parameters are not affected by the change of theoretical framework considered, showing how no degeneracies between standard and birefringence parameters are detected by WMAP+BICEP. Moreover the results obtained for $\alpha_1$, $\alpha_0$ and $\beta$ are comparable except for the time evolving angle which exhibits a sign opposite to the other two, due to the definition given in Section \ref{sec:theo}. Nevertheless, as already found in \cite{Barkats:2013jfa}, it must be noted that the combination WMAP+BICEP favors a non vanishing birefringence angle (see Fig.\ref{fig:monodim}, whether 
it produces time-varying, constant or late time rotation, at $\approx2.5\ \sigma$. \\

\begin{table}[!htb]
\centering
\begin{tabular}{|c|c|c|c|}
\hline
& $\beta$ rotation & $\alpha_0$ rotation & $\alpha_1$ rotation \\
\hline
Parameter & & &\\
\hline
$\Omega_bh^2$       &   $0.0228\pm0.0005$      &   $0.0228\pm0.0005$   & $0.0228\pm0.0005$\\
$\Omega_ch^2$       &   $0.112\pm0.004$        &   $0.112\pm0.004$     & $0.112\pm0.004$\\
$\theta$            &   $1.040\pm0.002$        &   $1.040\pm0.002$     & $1.040\pm0.002$\\
$\tau$              &   $0.089\pm0.01$         &   $0.089\pm0.01$      & $0.089\pm0.01$\\
$n_s$               &   $0.973\pm0.01$         &   $0.974\pm0.01$      & $0.974\pm0.01$\\
$\log(10^{10}A_s)$  &   $3.17\pm0.04$          &   $3.17\pm0.04$       & $3.17\pm0.04$\\
$\alpha_1$          &    $-$         &   $-$                 & $1.79\pm0.74$   \\
$\alpha_0$          &   $-$                    &   $-1.71\pm0.69$      & $-$  \\
$\beta$             &   $-1.77\pm0.71$                   &   $-$                 & $-$  \\
$H_0$               &   $70.9\pm2.1$           &   $70.9\pm2.2$        & $70.9\pm2.2$\\
\hline
\end{tabular}
\caption{Marginalized mean values and $68 \%$ c.l. errors on cosmological parameters using WMAP+BICEP data.}
\label{tab:resWMAP}
\end{table}

\subsection{Forecasted Results}
Once the best fit values for WMAP+BICEP are obtained in the three different cosmologies, we use them as the fiducial cosmology to forecast future constraints, achievable by the combination of Planck and PolarBear surveys. We investigate if the combination of these two experiments will be able to significantly distinguish the assumed cosmology from one where no rotation, of any kind, is present.\\

The first datasets we produce assume as fiducial cosmology the best fit for the early-time rotation with a constant angle $\alpha_0$. This is fitted by three different cosmologies: a varying $\alpha_0$ cosmology, to quantify the improvement brought by Planck+PolarBear on $\alpha_0$ constraints, a varying late time rotation angle $\beta$ cosmology, to investigate the possibility to distinguish the two rotation mechanisms with this combination of surveys, and a standard $\Lambda$CDM cosmology to quantify the possible bias brought on standard parameters by wrongly assuming a vanishing rotation angle.\\
In Table~\ref{tab:resPkPba0} we report the results obtained by combining Planck and PolarBear forecasted datasets and it is possible to notice how the improvement in sensitivity due to these surveys will allow to tighten the constraints on $\alpha_0$ and $\beta$ and  therefore, if WMAP+BICEP best fit values are confirmed, to rule out the possibility of a vanishing birefringence angle, both for the early and late time rotations.\\

\begin{table*}[!htb]
\centering
\begin{tabular}{|c|c|c|c|}
\hline
& $\alpha_0$ rotation & $\beta$ rotation & no rotation\\
\hline
Parameter & & & \\
\hline
$\Omega_bh^2$       &   $0.02279\pm0.00008$  & $0.02280\pm0.00008$   & $0.02242\pm0.00008$ \\  
$\Omega_ch^2$       &   $0.1123\pm0.0007$    & $0.1121\pm0.0007$     & $0.1192\pm0.0008$ \\ 
$\theta$            &   $1.0400\pm0.0002$    & $1.040\pm0.0002$      & $1.040\pm0.0002$ \\ 
$\tau$              &   $0.089\pm0.003$      & $0.892\pm0.003$       & $0.0947\pm0.004$ \\  
$n_s$               &   $0.974\pm0.002$      & $0.974\pm0.002$       & $0.961\pm0.002$ \\  
$\log(10^{10}A_s)$  &   $3.17\pm0.01$        & $3.17\pm0.01$         & $3.24\pm0.01$ \\   
$\alpha_0$          &   $-1.71\pm0.01$       & $-$                   & $-$ \\     
$\beta$             &   $-$                  & $-1.74\pm0.01$        & $-$ \\     
$H_0$               &   $70.8\pm0.3$         & $70.9\pm0.3$          & $67.8\pm0.3$ \\     
\hline
\end{tabular}
\caption{Marginalized mean values and $68 \%$ c.l. errors on cosmological parameters using Planck+PolarBear forecasted data in the $\alpha_0$, $\beta$ and no rotation analysis. The fiducial model used to build the simulated dataset is based on the WMAP+BICEP analysis including $\alpha_0$ (i.e. early time rotation), second column of Table \ref{tab:resWMAP}}
\label{tab:resPkPba0}
\end{table*}

\begin{figure}[!htb]
\begin{center}
\hspace*{-1cm}
 \includegraphics[width=8cm]{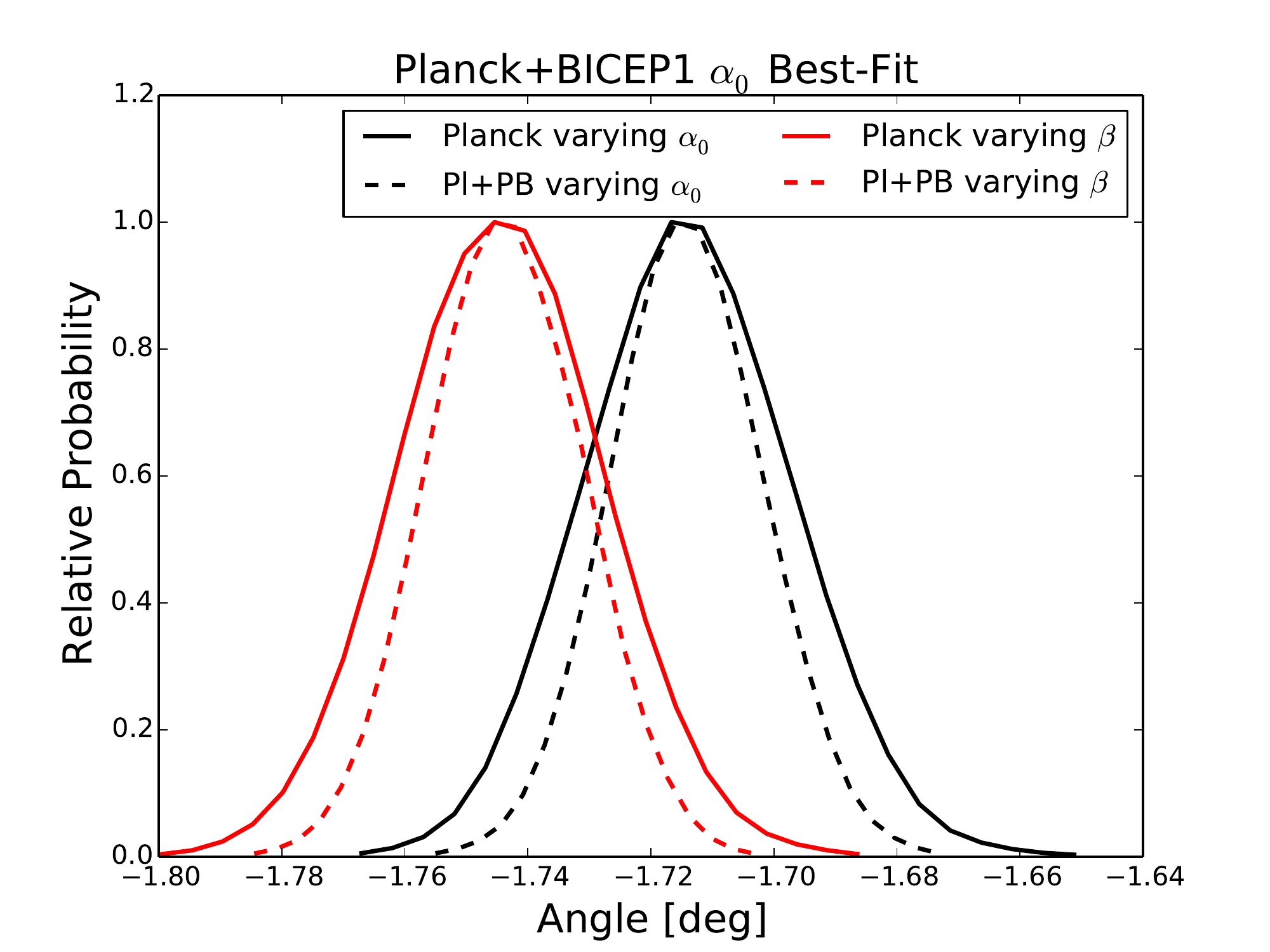} 
\caption{Posterior distribution for early-type and late-time birefringence parameters $\alpha_0$ (black lines) and $\beta$ (red lines) using Planck (solid lines) and Planck+PolarBear (dashed lines) forecasted datasets, with a fiducial cosmology equivalent to the second column of Table \ref{tab:resWMAP}.}
\label{fig:monodim_a0_beta_mocka0}
\end{center}
\end{figure}

\begin{figure}[!h]
\begin{center}
\hspace*{-1cm}
 \includegraphics[width=12cm]{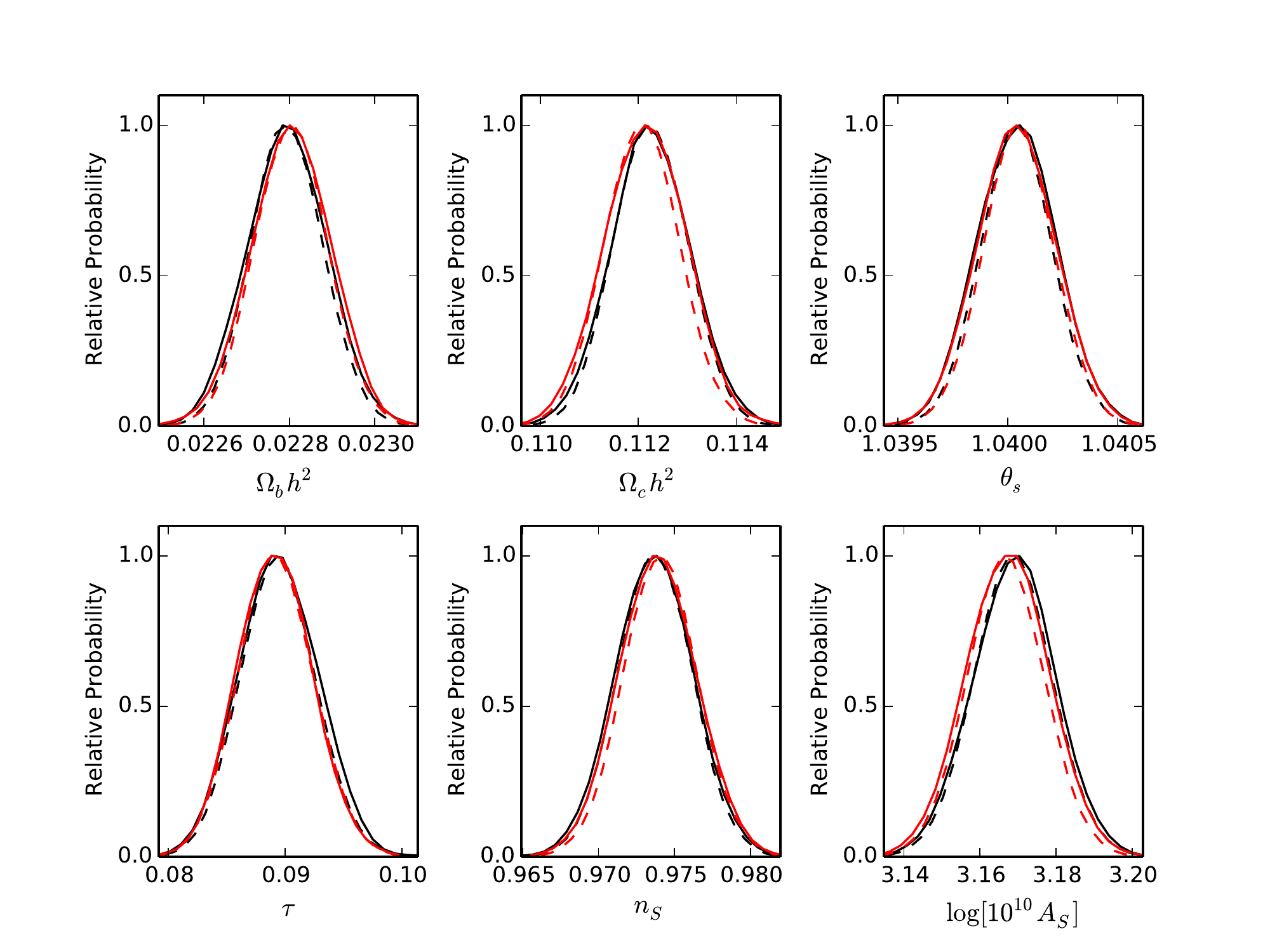} 
\caption{Posterior distribution for several standard cosmological parameters using Planck (solid lines) and Planck+PolarBear (dashed lines) forecasted datasets, with a fiducial cosmology equivalent to the second column of Table \ref{tab:resWMAP}, analyzed with a varying $\alpha_0$ (black lines) and with a varying $\beta$ (red lines). We see that even with Planck+PolarBear sensitivity there is no detectable effect on the standard cosmological parameters}
\label{fig:monodim_a0_beta_mocka0_params}
\end{center}
\end{figure}

\begin{figure}[!h]
\begin{center}
\hspace*{-1cm}
\begin{tabular}{cc}
 \includegraphics[width=12cm]{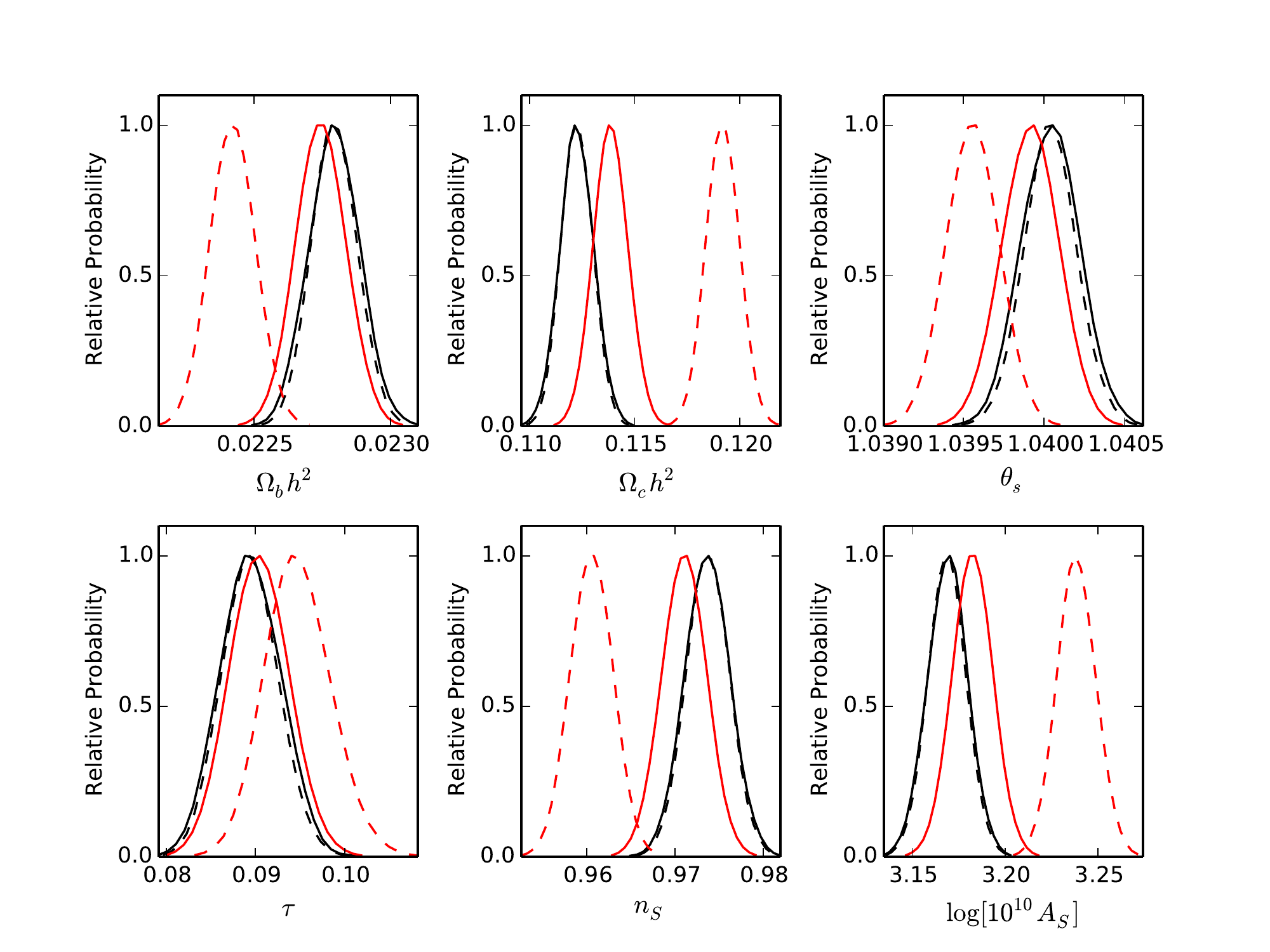}
 \end{tabular}
\caption{Posterior distribution for several standard cosmological parameters using Planck (solid lines) and Planck+PolarBear (dashed lines) forecasted datasets, with a fiducial cosmology equivalent to the second column of Table \ref{tab:resWMAP}, analyzed with a varying $\alpha_0$ (black lines) and with no rotation cosmology (red lines). We see that with Planck sensitivity the effect of disregarding the rotation is significant on the standard cosmological parameters, and it becomes catastrophic with Planck+PolarBear sensitivity.}
\label{fig:monodim_a0_standard_mocka0}
\end{center}
\end{figure}

When fitting the considered datasets with a varying $\beta$ no shift is detected in the recovered values of the standard cosmological parameters, as can be seen in Fig.\ref{fig:monodim_a0_beta_mocka0_params}, highlighting the fact that the varying $\beta$ cosmology is able to reproduce the fiducial one. However the recovered amount of rotation is significantly different if different theoretical models for the rotation are used in the fit, see Figure \ref{fig:monodim_a0_beta_mocka0}. As the $\beta$ rotation angle can be thought, due to its properties, as a systematic effect arising from calibration errors \cite{malapola}, the results we show highlight once more the necessity to minimize these errors as a poor calibration could lead to false detection of physical rotation effects.\\
When the datasets are instead analyzed with a standard $\Lambda$CDM cosmology without any  polarization rotation we obtain a shift of the order of several standard deviations in standard cosmological parameters; this effect, clearly visible in Fig.\ref{fig:monodim_a0_standard_mocka0}, arises from the fact that the theoretical spectra produced assuming a vanishing rotation angle are not able to reproduce the simulated datasets polarization spectra (see Fig. \ref{fig:spectra}).\\

As a further step we inquire whether Planck+PolarBear can distinguish between an early times constant rotation and a time evolving one; in order to do this, we use the WMAP+BICEP best fit value, obtained assuming a time evolving rotation, as fiducial cosmology and we analyze the obtained datasets again with three different cosmology, i.e. with $\alpha_0$ or $\alpha_1$ free to vary and with standard $\Lambda$CDM.\\
In Table~\ref{tab:resPkPba1} we report the results obtained combining Planck and PolarBear forecasted datasets and again we notice how the constraining power of Planck+PolarBear allows to rule out the vanishing birefringence angle cosmology.\\

\begin{table*}[!htb]
\centering
\begin{tabular}{|c|c|c|c|}
\hline
& $\alpha_1$ rotation & $\alpha_0$ rotation & no rotation\\
\hline
Parameter & & & \\
\hline
$\Omega_bh^2$       &   $0.02281\pm0.00008$   & $0.02281\pm0.00008$  & $0.02245\pm0.00008$ \\
$\Omega_ch^2$       &   $0.1122\pm0.0007$     & $0.1122\pm0.0007$    & $0.1191\pm0.0008$ \\
$\theta$            &   $1.040\pm0.0002$      & $1.040\pm0.0002$     & $1.040\pm0.0002$ \\
$\tau$              &   $0.089\pm0.003$       & $0.089\pm0.003$      & $0.094\pm0.004$ \\
$n_s$               &   $0.974\pm0.002$       & $0.974\pm0.002$      & $0.961\pm0.002$ \\
$\log(10^{10}A_s)$  &   $3.17\pm0.01$         & $3.17\pm0.01$        & $3.24\pm0.01$ \\
$\alpha_1$          &   $1.79\pm0.01$         & $-$                  & $-$ \\
$\alpha_0$          &   $-$                   & $-1.69\pm0.01$       & $-$ \\
$H_0$               &   $70.8\pm0.3$          & $70.9\pm0.3$         & $67.9\pm0.3$ \\
\hline
\end{tabular}
\caption{Marginalized mean values and $68 \%$ c.l. errors on cosmological parameters using Planck+PolarBear forecasted data in the $\alpha_1$, $\alpha_0$ and no rotation analysis. The fiducial model used to build the simulated dataset is based on the WMAP+BICEP analysis including $\alpha_1$ (i.e. time evolving rotation), third column of Table \ref{tab:resWMAP}}
\label{tab:resPkPba1}
\end{table*}

\begin{figure}[!h]
\begin{center}
\hspace*{-1cm}
\begin{tabular}{cc}
 \includegraphics[width=12cm]{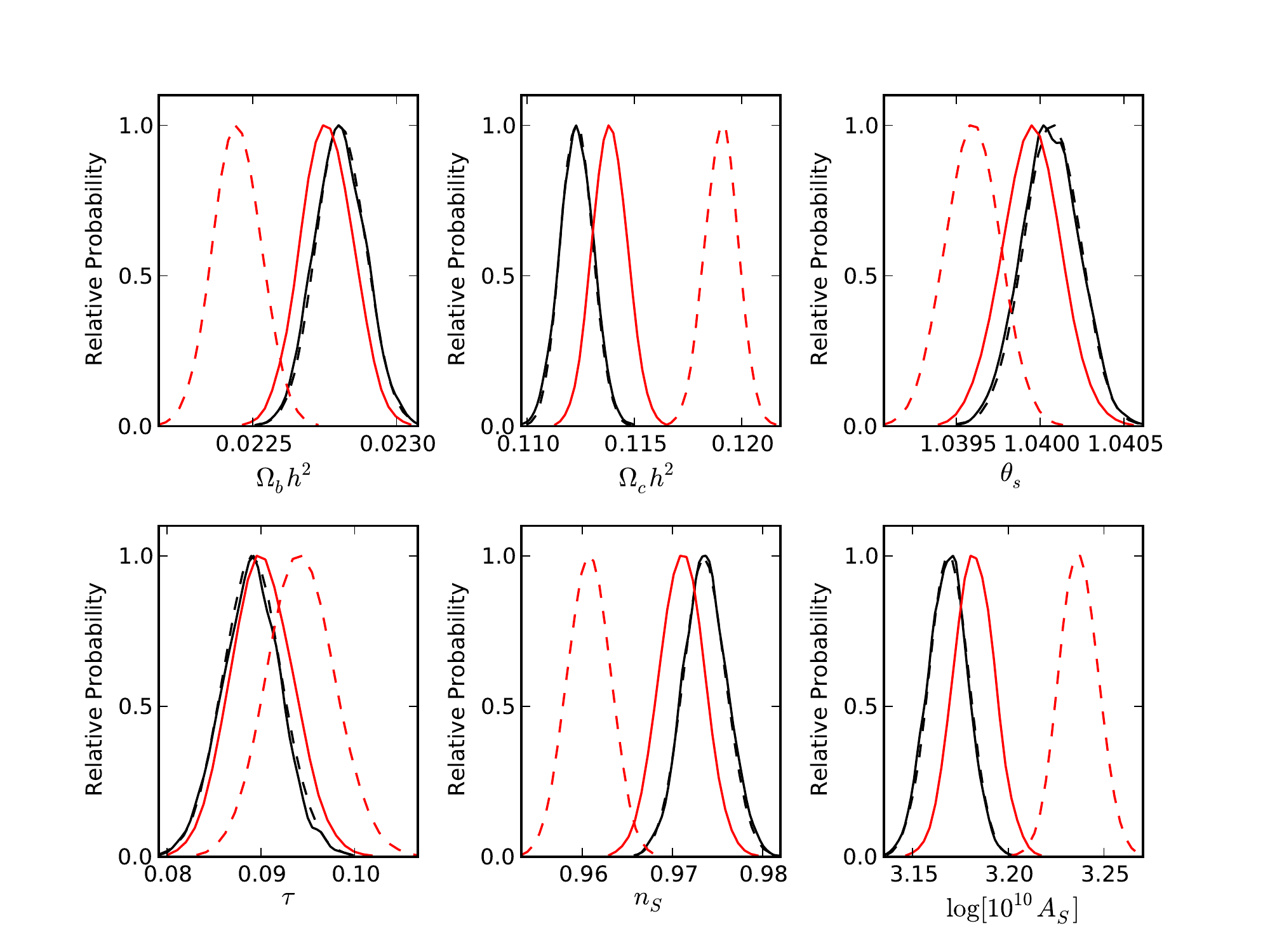}
 \end{tabular}
\caption{Posterior probability distribution for the standard cosmological parameters using Planck (solid lines) and Planck+PolarBear (dashed lines) forecasted datasets, with a fiducial cosmology equivalent to the third column of Table \ref{tab:resWMAP}, analyzed with a varying $\alpha_1$ (black lines) and with no rotation cosmology (red lines). We see that with Planck sensitivity the effect of disregarding the rotation is significant on the standard cosmological parameters, and it becomes catastrophic with Planck+PolarBear sensitivity.}
\label{fig:monodim_a1_standard_mocka1}
\end{center}
\end{figure}

As in the previous analysis, we notice also in this case how the combination of Planck and PolarBear is unable to distinguish the two rotation mechanisms while we again find a shift of a few $\sigma$ order of magnitude when the datasets are analyzed assuming a vanishing rotation angle (see Fig. \ref{fig:monodim_a1_standard_mocka1}).\\

\begin{figure}[!h]
\begin{center}
\hspace*{-1cm}
 \includegraphics[width=12cm]{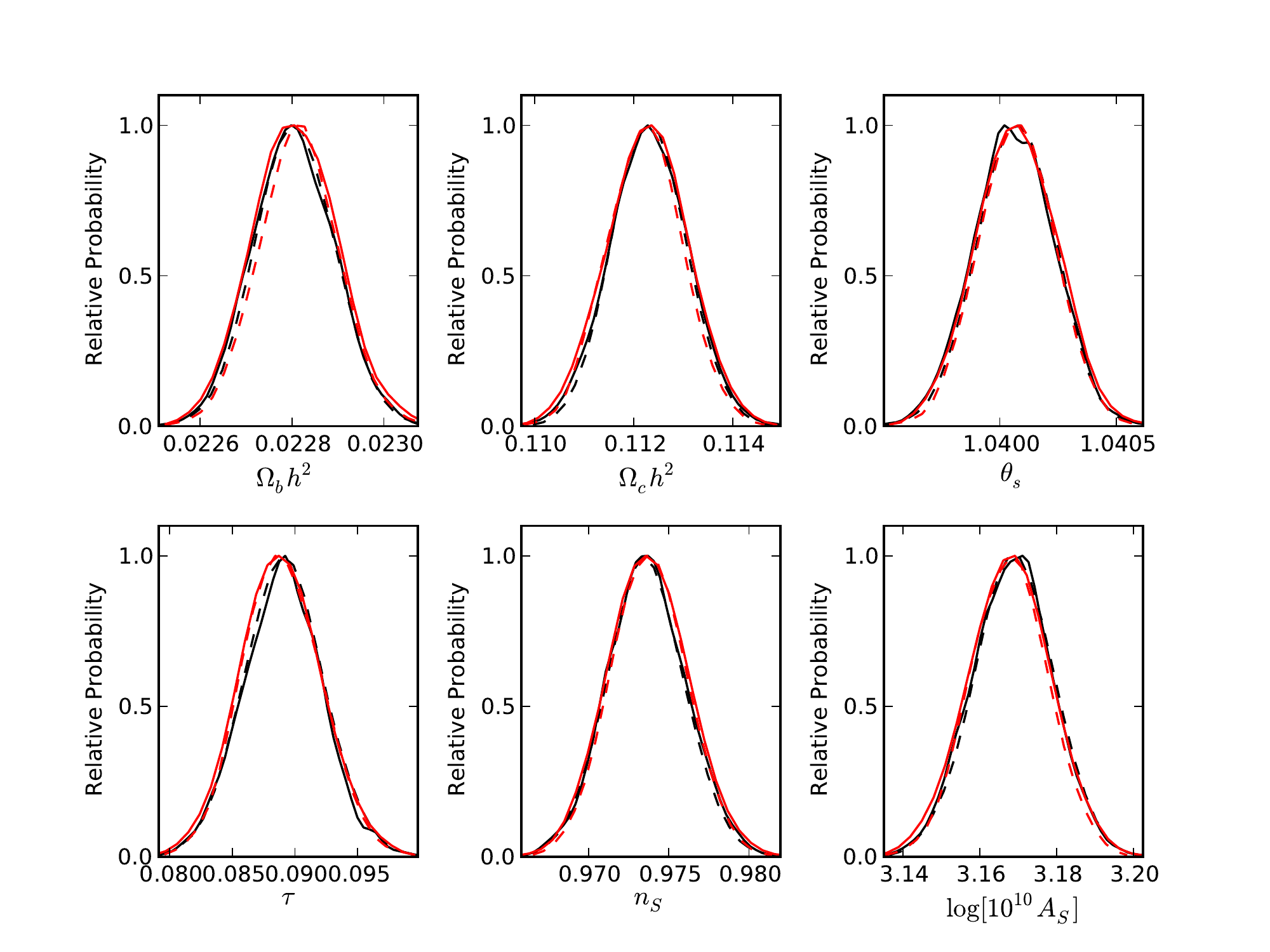} 
\caption{Posterior probability distribution for the standard cosmological parameters using Planck (solid lines) and Planck+PolarBear (dashed lines) forecasted datasets, with a fiducial cosmology equivalent to the third column of Table \ref{tab:resWMAP}, analyzed with a varying $\alpha_1$ (black lines) and with a varying $\alpha_0$ (red lines). We see that even with Planck+PolarBear sensitivity there is no detectable effect on the standard cosmological parameters}
\label{fig:monodim_a1_a0_mocka1_params}
\end{center}
\end{figure}

\subsection{Lensing Degeneracies}

As already stated in Section \ref{sec:theo} and shown in Fig.~\ref{fig:spectra}, one of the effects of a non vanishing birefringence angle is to shift power from the EE to the BB modes. Another source of mixing between EE and BB modes is brought by CMB lensing which produces a non vanishing BB spectrum even if a birefringence angle is not present (see Fig.~\ref{fig:lensdeg}). In order to understand if degeneracies between the two effects exist, we analyze the simulated Planck+PolarBear datasets, obtained with time evolving rotation as fiducial cosmology, assuming no rotation is present, but allowing for a varying lensing amplitude $A_L$.\\

\begin{table*}[!htb]
\centering
\begin{tabular}{|c|c|c|c|c|c|}
\hline
& $\alpha_0$ rotation & no rotation &$\Delta/\sigma$ & no rotation$+A_L$ &$\Delta/\sigma$\\
\hline
Parameter & & & & &\\
\hline
$\Omega_bh^2$       &   $0.02279\pm0.00008$   & $0.02242\pm0.00008$ & $0.5$  &$0.02278\pm0.00009$ & $0.1$\\
$\Omega_ch^2$       &   $0.1123\pm0.0007$     & $0.1192\pm0.0008$   & $8.6$  &$0.1136\pm0.0009$   & $1.4$\\
$\theta$            &   $1.040\pm0.0002$      & $1.040\pm0.0002$    & $0.0$  &$1.040\pm0.0002$    & $0.0$\\
$\tau$              &   $0.089\pm0.003$       & $0.095\pm0.004$     & $1.5$  &$0.088\pm0.003$     & $0.3$\\
$n_s$               &   $0.974\pm0.002$       & $0.961\pm0.002$     & $6.5$  &$0.970\pm0.002$     & $2.0$\\
$\log(10^{10}A_s)$  &   $3.17\pm0.01$         & $3.24\pm0.01$       & $7.0$  &$3.18\pm0.01$       & $1.0$\\
$\alpha_1$          &   $-1.71\pm0.01$         & $-$                & $-$    &$-$                 & $-$  \\
$A_L$               &   $-$                   & $-$                 & $-$    &$1.29\pm0.03$       & $-$  \\
$H_0$               &   $70.8\pm0.3$          & $67.8\pm0.3$        & $10$   &$70.3\pm0.4$        & $1.3$\\
\hline
\end{tabular}
\caption{Marginalized best fit values and $68 \%$ c.l. errors on cosmological parameters using Planck+PolarBear simulated data (assumed fiducial cosmology obtained from third column of \ref{tab:resWMAP}) analyzed with $\alpha_0$ rotation, no rotation and vanishing rotation angle when a variation of the $A_L$ parameter is allowed. Third and fifth  columns report the shift with respect to the varying $\alpha_0$ best fit in unity of $\sigma$, for the no rotation and free $A_L$ cases respectively. }
\label{tab:resAl}
\end{table*}

\begin{table*}[!htb]
\centering
\begin{tabular}{|c|c|c|c|c|c|}
\hline
& $\alpha_0$ rotation & no rotation &$\Delta/\sigma$ & no rotation$+A_L$ &$\Delta/\sigma$\\
\hline
Parameter & & & & &\\
\hline
$\Omega_bh^2$       &   $0.02279\pm0.00009$   & $0.02274\pm0.00009$ & $0.5$  &$0.02283\pm0.00009$ & $0.4$\\
$\Omega_ch^2$       &   $0.1123\pm0.0008$     & $0.1138\pm0.0008$   & $1.9$  &$0.1126\pm0.0009$   & $0.4$\\
$\theta$            &   $1.040\pm0.0002$      & $1.040\pm0.0002$    & $0.0$  &$1.040\pm0.0002$    & $0.0$\\
$\tau$              &   $0.089\pm0.003$       & $0.090\pm0.003$     & $0.3$  &$0.089\pm0.003$     & $0.0$\\
$n_s$               &   $0.974\pm0.002$       & $0.971\pm0.002$     & $1.5$  &$0.973\pm0.002$     & $0.5$\\
$\log(10^{10}A_s)$  &   $3.17\pm0.01$         & $3.18\pm0.01$       & $1.0$  &$3.17\pm0.01$       & $0.0$\\
$\alpha_1$          &   $-1.71\pm0.02$         & $-$                & $-$    &$-$                 & $-$  \\
$A_L$               &   $-$                   & $-$                 & $-$    &$1.09\pm0.03$       & $-$  \\
$H_0$               &   $70.8\pm0.4$          & $70.1\pm0.4$        & $2.0$   &$70.3\pm0.4$        & $0.3$\\
\hline
\end{tabular}
\caption{Marginalized best fit values and $68 \%$ c.l. errors on cosmological parameters using Planck simulated data (assumed fiducial cosmology obtained from third column of \ref{tab:resWMAP}) analyzed with $\alpha_0$ rotation, no rotation and vanishing rotation angle when a variation of the $A_L$ parameter is allowed. Third and fifth  columns report the shift with respect to the varying $\alpha_0$ best fit in unity of $\sigma$, for the no rotation and free $A_L$ cases respectively. }
\label{tab:resAlplanck}
\end{table*}

\begin{figure}[!h]
\begin{center}
\hspace*{-1cm}
 \includegraphics[width=12cm]{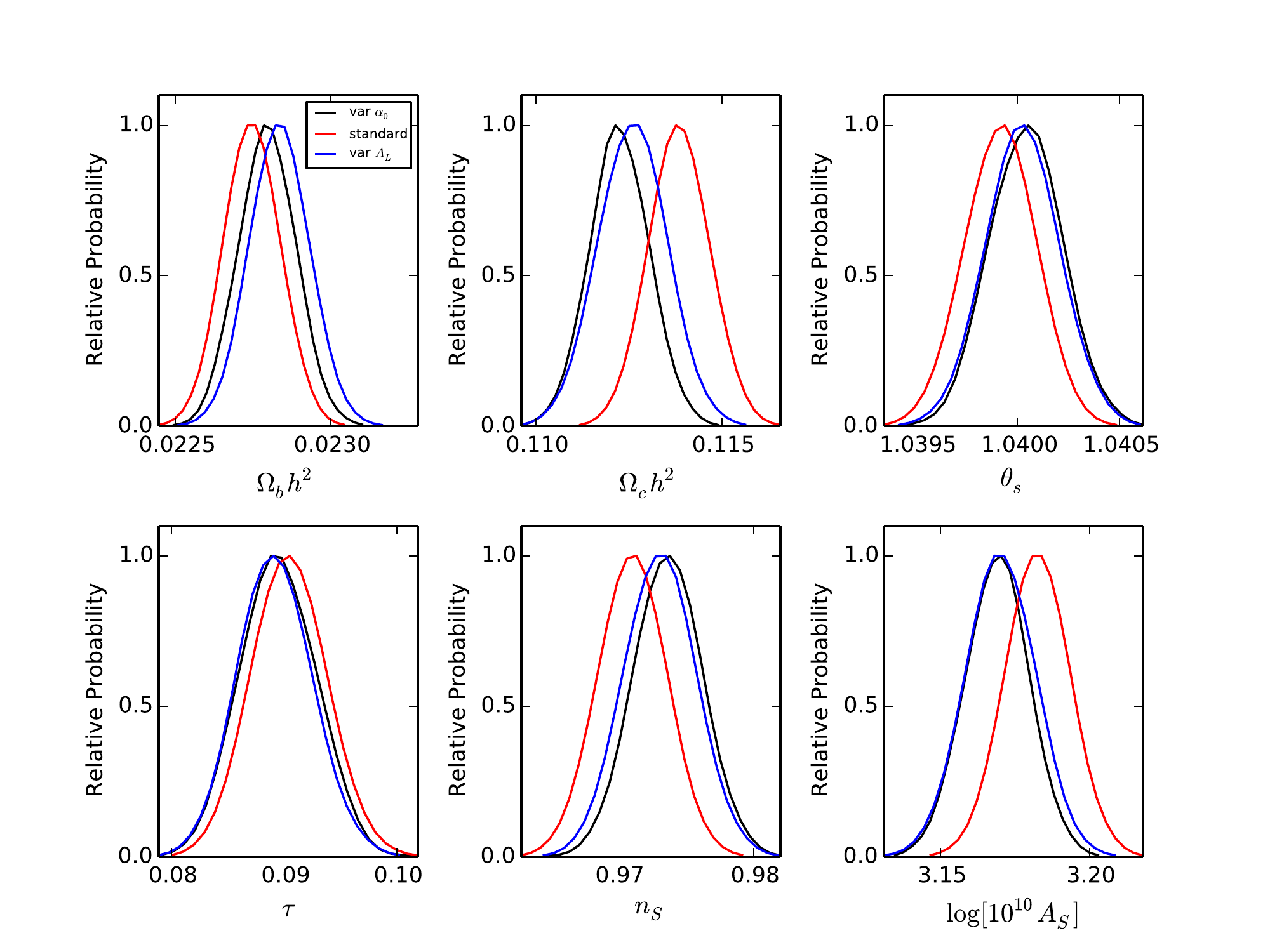}  
\caption{Posterior distribution for the parameters using Planck forecasted datasets, with a fiducial cosmology equivalent to the first column of Table \ref{tab:resWMAP}, analyzed with a varying $\alpha_0$ (black lines), with a varying $A_L$ (blue lines) and with only standard cosmological parameters (red lines). One sees that $A_L$ can compensate for the effects of birefringence quite well in experiments with Planck sensitivity}
\label{fig:monodim_pkAL}
\end{center}
\end{figure}

\begin{figure}[!h]
\begin{center}
\hspace*{-1cm}
 \includegraphics[width=12cm]{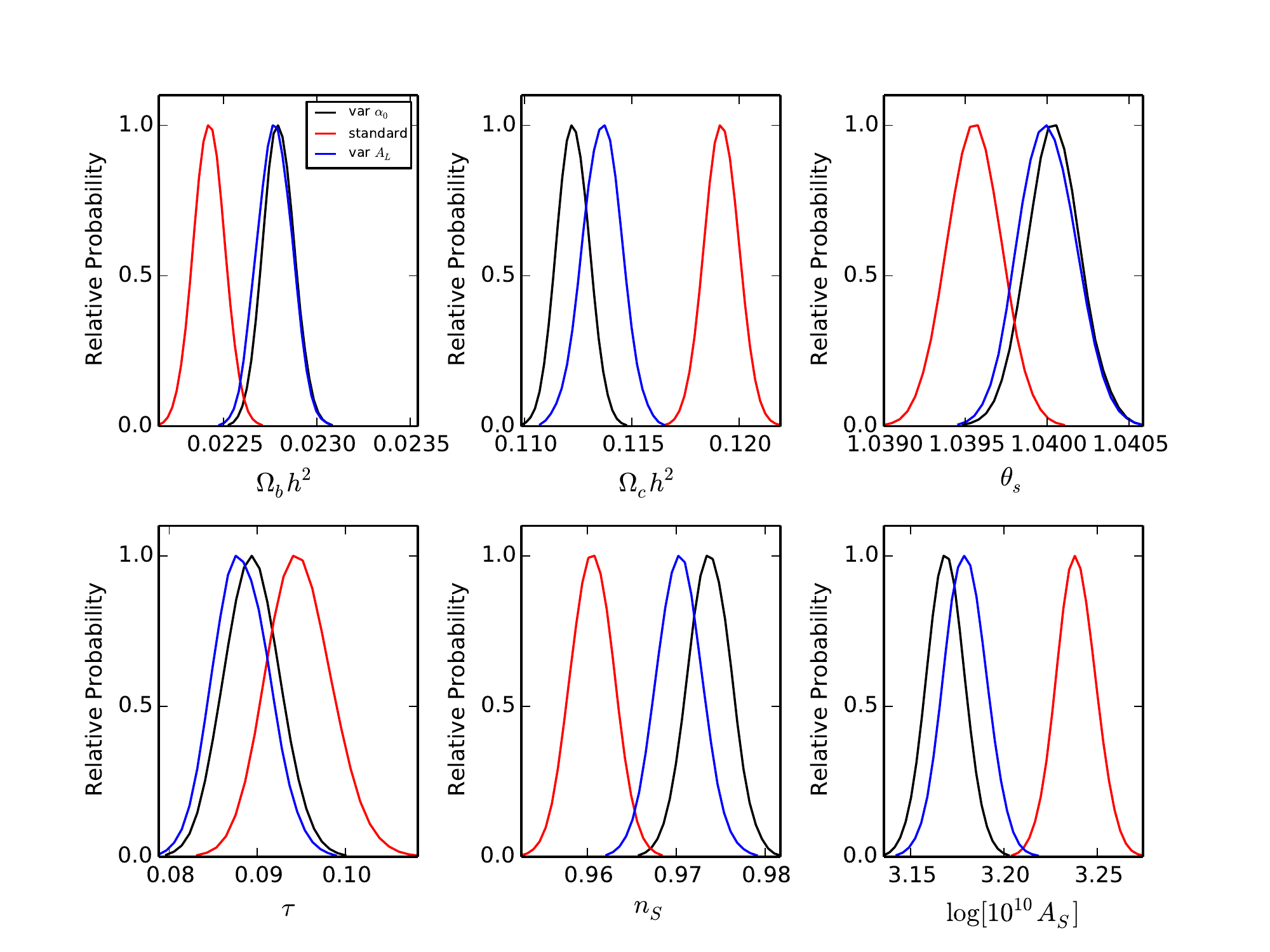}  
\caption{Posterior distribution for the parameters using Planck+PolarBear forecasted datasets, with a fiducial cosmology equivalent to the first column of Table \ref{tab:resWMAP}, analyzed with a varying $\alpha_0$ (black lines), with a varying $A_L$ (blue lines) and with only standard cosmological parameters (red lines). One sees that with the increased sensitivity provided by PolarBear one gets wrong values for the cosmological parameters if birefringence is disregarded. Moreover some parameters are sensitive to the different effects of $A_L$ and birefringence, so that $A_L$ can not always compensate for the effects of birefringence.}
\label{fig:monodim_pkpbAL}
\end{center}
\end{figure}

In Table \ref{tab:resAl} we report the obtained parameters and it is possible to notice (see also in Figs. \ref{fig:monodim_pkAL}-\ref{fig:monodim_pkpbAL}) how the presence of a varying $A_L$ mitigates the shift produced assuming the wrong cosmological model. This is due to the fact that raising the value of $A_L$ increases the power transfer from $EE$ to $BB$ modes and therefore can partially account for enhancement in the $BB$ spectrum produced by birefringence when the datasets are analyzed assuming vanishing rotation angles. It is interesting to point out how the shift in the standard parameters is mitigated at the price of a non standard value of $A_L$; this means that neglecting the presence of a birefringence rotation can possibly lead to a false detection of $A_L>1$ when analyzing data with Planck+PolarBear sensitivity.\\ 
We found similar, but less significant, results for Planck alone also. The recovered value of $A_L$ is $2.5\sigma$ away from the standard value and the other parameters are recovered within $1\sigma$.\\
However, we point out that, should this effect being observed in upcoming data, a practical way to distinguish between a non vanishing $\alpha$ cosmology and an enhanced lensing amplitude would lie in the analysis of the $\chi^2$ values obtained when fitting the data; while partially degenerate, in fact, the two cosmologies produce different effect on the CMB spectra (as can be seen in Fig.\ref{fig:lensdeg}) and therefore will produce different values of the $\chi^2$ when used to analyze datasets. This implies that Bayesian model selection techniques can be used to quantitatively understand which of the two theoretical models would be preferred by observations.\\

\section{Conclusions}
\label{sec:conc}

In this paper we investigated the possibility for current and upcoming data to measure the effect of a rotation of CMB photons polarization direction. This kind of effect can be ascribed either to calibration issues or to departures from the standard $\Lambda$CDM theory and it produces a peculiar mixing of E and B modes. When the effect is due to new physics, it is called birefringence, and its effects are different from the ones of a calibration-driven mixing because the rotation of polarization accumulates starting from early times, thus it affects CMB spectra before the E-B mixing due to gravitational lensing takes place.\\
We have considered three scenarios. The first one is a late-time E-B mixing (possibly due to calibration issues), parameterized by the polarization rotation angle $\beta$. This is the mixing more widely considered in the literature and is applied on the final CMB spectra, already affected by lensing . The second scenario is an early-time E-B mixing, which is a constant mixing happening before lensing, parameterized by the polarization rotation angle $\alpha_0$.
This is a first approximation of a more physically sound model in which the  polarization rotation is due to non-standard physics and accumulates during photon propagation. This was actually the last scenario we considered: a mixing of E and B modes that accumulates during CMB photon propagation with a linear time dependence. It is parameterized by the dimensionless quantity $\alpha_1$.
We analyzed WMAP CMB data alone and combined with BICEP data, finding that the sensitivity of these survey is not enough to distinguish among the three scenarios, but it is sufficient to show a hint for a non-vanishing value of either $\alpha_1$, $\alpha_0$ or $\beta$, pointing out how the combination of these datasets is able to detect the effect of polarization mixing in CMB spectra.\\
Furthermore, we used the parameters' values obtained with the latter analysis to forecast upcoming CMB data from Planck and PolarBear, in order to investigate the impact of these surveys on the possible detection of the three rotation scenarios and the possibility to distinguish between early and late time rotations and between constant and time-evolving rotation angles. We found, as expected, that the Planck+PolarBear analysis will narrow the constraints on rotation parameters with respect to WMAP+BICEP, possibly ruling out the vanishing rotation scenario, while it will not be able to distinguish among the three rotation mechanisms as, even assuming the wrong scenario, the theoretical spectra produced are still able to fit the fiducial cosmology.\\
Finally, we inquired about the possible degeneracy between birefringence and CMB lensing, as both physical phenomena produce a leaking from $E$ to $B$ modes; we found that the similar effect on CMB spectra leads to the possibility of partially mimicking the birefringence driven enhancement of standard $\Lambda$CDM B modes with a cosmology where a vanishing rotation is assumed, but a non standard amplitude of CMB lensing is allowed; this result shows how the sensitivity that will be achieved by Planck+PolarBear will be high enough to prompt the need for very accurate analysis of these effects as a false detection of a non standard lensing could be obtained if the birefringence effect is neglected.

\section*{AKNOWLEDGMENTS}
We would like to thank Carlo Baccigalupi and Alessandro Melchiorri for useful discussion.
MM acknowledges partial support from the INDARK INFN grant. GG was supported in part by a
grant from the John Templeton Foundation. 

\bibliographystyle{apsrev4-1}


\end{document}